\documentclass[sigconf,natbibnatbib=true,anonymous=false]{acmart}
\usepackage{graphicx} 
\usepackage{float} 
\usepackage{subfigure} 
\usepackage{algorithm} 
\usepackage{multirow} 
\usepackage{amsmath}
\usepackage{xcolor}
\usepackage{algpseudocode}
\usepackage{threeparttable} 
\usepackage{bm}
\usepackage{enumitem}
\usepackage[switch]{lineno}

\renewcommand{\algorithmicrequire}{\textbf{Input:}}  
\renewcommand{\algorithmicensure}{\textbf{Output:}} 
\renewcommand\footnotetextcopyrightpermission[1]{} 

\AtBeginDocument{%
  \providecommand\BibTeX{{%
    \normalfont B\kern-0.5em{\scshape i\kern-0.25em b}\kern-0.8em\TeX}}}

\setcopyright{acmcopyright}
\copyrightyear{2018}
\acmYear{2018}
\acmDOI{10.1145/1122445.1122456}

\begin{document}

\title{NCS4CVR: Neuron-Connection Sharing for Multi-Task Learning in Video Conversion Rate Prediction}

\author{Xuanji Xiao}
\authornote{Both authors contributed equally to this research.}
\authornote{Xuanji Xiao is the Corresponding author.}
\orcid{0000-0002-0499-8838}
\affiliation{%
  \institution{Tencent Inc}
  \streetaddress{Xigema building,Zhichun Road, Haidian qu}
  \city{Beijing}
  \country{China}
  \postcode{100080}
}
\email{stevexiao@tencent.com}

\author{Huabin Chen}
\authornotemark[1]
\affiliation{%
  \institution{Tencent Inc}
   \city{Beijing}
  \country{China}
}
\email{490269628c@gmail.com}

\author{Yuzhen Liu}
\affiliation{%
  \institution{Tencent Inc}
  \streetaddress{Xigema building,Zhichun Road, Haidian qu}
  \city{Beijing}
  \country{China}}
\email{yzhenliu@tencent.com}

\author{Xing Yao}
\affiliation{%
   \institution{Tencent Inc}
  \streetaddress{Xigema building,Zhichun Road, Haidian qu}
  \city{Beijing}
  \country{China}}
\email{jessieyao@tencent.com}

\author{Chaosheng Fan}
\affiliation{%
   \institution{Tencent Inc}
  \streetaddress{Xigema building,Zhichun Road, Haidian qu}
  \city{Beijing}
  \country{China}}
\email{atlasfan@tencent.com}

\author{Pei Liu}
\affiliation{%
   \institution{Tencent Inc}
  \streetaddress{Xigema building,Zhichun Road, Haidian qu}
  \city{Beijing}
  \country{China}}
\email{alexpliu@tencent.com}

\renewcommand{\shortauthors}{Anonymous Author, et al.}

\begin{abstract}
Click-through rate (CTR) and post-click conversion rate (CVR) predictions are two fundamental modules in industrial ranking systems such as recommender systems, advertising, and search engines.
 Since CVR involves much fewer samples than CTR (known as the CVR data sparsity problem), most of the existing works try to leverage CTR\&CVR multi-task learning to improve CVR performance.
 However, typical coarse-grained sub-network/layer sharing methods may introduce conflicts and lead to performance degradation, since not every neuron or neuron connection in one layer should be shared between CVR and CTR tasks.
 This is because users may have different fine-grained content feature preferences between deep consumption and click behavior, represented by CVR and CTR, respectively.
 To address this sharing\&conflict problem, we propose a novel multi-task CVR modeling scheme with neuron-connection level sharing named NCS4CVR, which can automatically and flexibly learn which neuron weights are shared or not shared without artificial experience.
Compared with previous layer-level sharing methods, this is the first time that a fine-grained CTR\&CVR sharing method at the neuron connection level is proposed, which is a research paradigm shift in the sharing level. 
Both offline and online experiments demonstrate that our method outperforms both the single-task model and the layer-level sharing model.
Our proposed method has now been successfully deployed in an industry video recommender system serving major traffic.
 \footnote{The source code is available. https://github.com/aunusualman/LT4REC.}
\end{abstract}

\maketitle

\section{Introduction}

\begin{figure}[tbp] 
\centering 
\includegraphics[width=\columnwidth]{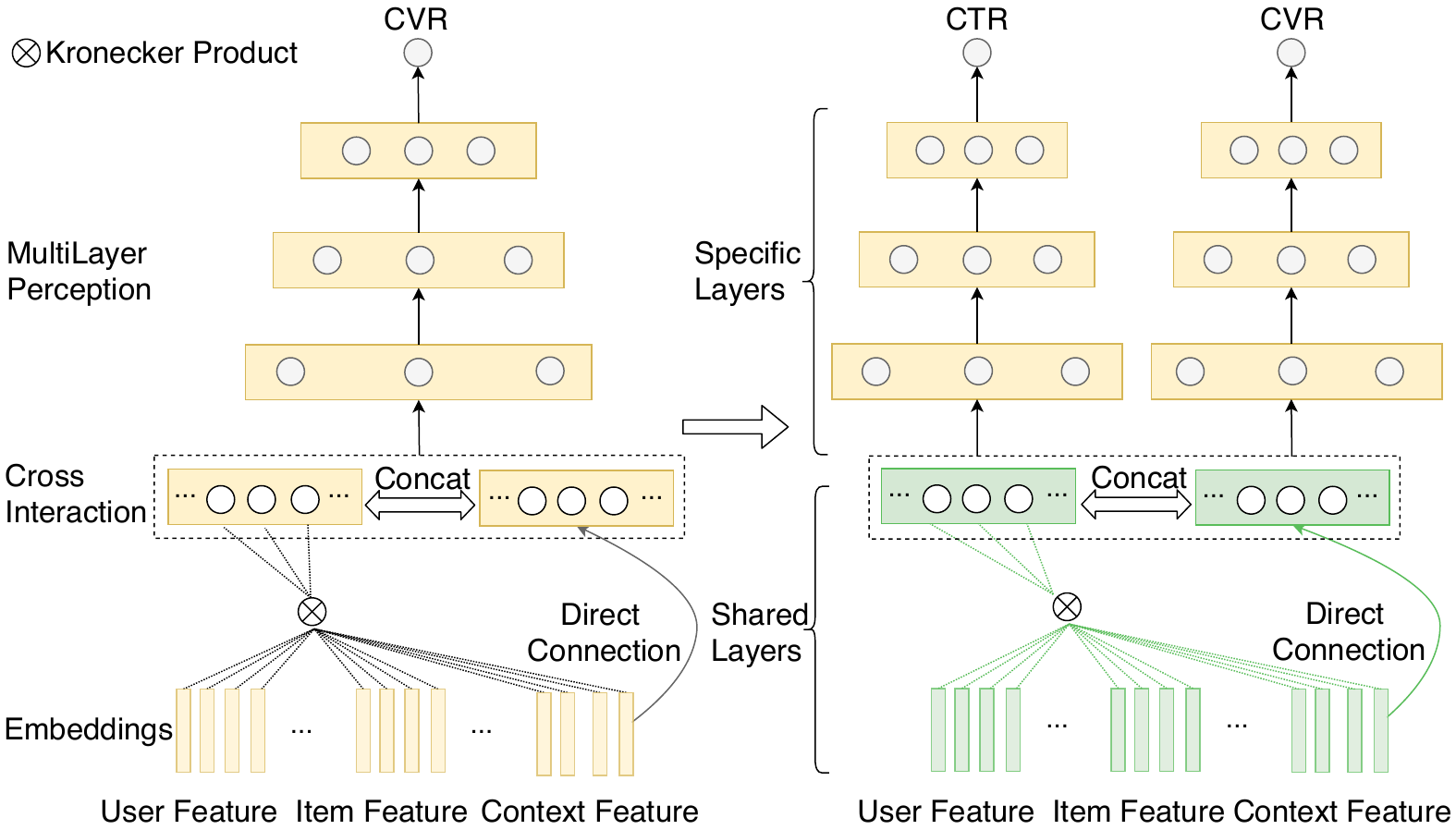}
\vspace{-18pt}
\caption{Architecture overview of the typical single task CVR model (left) and the classic layer level sharing CVR model (right). 
} 
\label{fig.main2} 
\vspace{-15pt}
\end{figure}

 Post-click conversion rate (CVR) prediction plays a key role across industrial ranking systems, such as recommender systems (RS) \cite{ma2018entire,wen2020entire,tang2020progressive}, online advertising \cite{zhu2017optimized}, and search engines \cite{zhang2020towards}. In video recommendation, an RS first recalls a large number of related videos and then ranks and exposes them to the users according to several ranking metrics, e.g., click-through rate (CTR) or CVR. Finally, users may click on (CTR sample) and eventually view the videos (CVR sample). For network media contents the CVR indicator is continuous, $label\in\left[0,1\right]$, e.g., if someone watched 2 minutes of a video with a length of 5 minutes then $CVR=0.4$.
 The number of CVR samples equals the video click number, which are the quantity of the positive samples of the CTR task. Thus the main challenge of CVR modeling is the data sparsity involving much fewer samples than CTR, which imposes a challenging model fitting process resulting in a poor generalization ability. A feasible solution is to use the CTR task to improve the performance of the CVR task with multi-task learning (MTL) as Fig.\ref{fig.main2}. 

Many studies have tried to optimize the CVR task by taking advantage of more CTR training samples via multi-task learning \cite{ma2018entire,wen2020entire,zhao2019recommending,ma2018modeling,ma2019snr,tang2020progressive}. 
ESMM series (ESMM, ESMM2 \cite{ma2018entire,wen2020entire}) uses click signals (CTR) and post-click signals (like rate,collection rate, etc.) to tackle the conversion sample sparsity. This strategy handles the problem by using feature-embedding layer sharing, which relies on manual experience.  
MMOE series (MMOE, SNR, PLE \cite{ma2018modeling,zhao2019recommending,ma2019snr,tang2020progressive}) model multi-task learning with a multi-gate mixture of expert sub-networks, which provide a limited degree of sharing at the granularity of the sub-network.
Each expert sub-network will learn a different representation, and then a task will use the weighted fusion of the learned representations to learn its final representation. 
As we have seen, the existing literature has provided a certain degree of sharing mechanism based on sub-networks or neural layers, so we collectively refer to these tasks as the layer level CVR sharing method. 

Despite their success, this coarse-grained layer-level sharing method often introduces sharing conflicts, especially when the tasks are loosely related. \cite{zhang2018overview}. 
MTL is proven to improve performance through information sharing between tasks \cite{tang2020progressive}.
However, not every neuron or neuron connection in a layer should be shared across both CVR and CTR tasks. 
For example, users may have a random click or a click for an exaggerated title or unreal cover image, and this behavior will be learned by specific neuron-connection weights in the CTR-CVR multi-task model through the CTR task. Still, this learned representation should not be shared by the CVR task because the CVR score won't be high in this situation.

To address the above \textbf{sharing\&conflict} issue, we propose a novel neuron-connection level sharing multi-task CVR modeling based on the lottery ticket theory \cite{frankle2018the}, learning which neuron connection to be shared automatically without artificial experience. The lottery ticket theory recently attracts great attention in neural network pruning \cite{malach2020proving,2020Learning}. This method states that a randomly initialized network contains a small sub-network (winning tickets) such that, when trained in isolation, can compete with the performance of the original large network. In other words, each task has its own best sub-network structure and don't need a dense and large network for training. The lottery ticket is first applied to the field of network model compression for finding sparse, trainable neural networks. 
However, this neuron-connection level-sparse structure exploration can be used for MTL, with the hypothesis that some neural connections should be shared across different tasks while others should not in a multi-task neural network model (MTL winning tickets). In other words, different tasks can have their own networks while also sharing certain-node connections \cite{2020Learning}. 
We start with an over-parameterized deep learning recommendation model, called the base network, from which CTR and CVR extract their subnetworks. The two subnetworks are partially overlapped. CTR subnetwork and CVR subnetwork are trained alternately, during which each task only updates the neuron weights of its corresponding subnetwork (Algorithm 1). In this way, CTR and CVR can have their own networks while also sharing certain node connections.
Particularly, we may find the best sub-networks for CTR and CVR automatically, in which the overlapping part and the specific part correspond to sharing representation and conflict reduction, respectively (Fig.\ref{fig.main3}). 

Our main contributions can be summarized as: 

\begin{itemize}[leftmargin=*]
\item To the best of our knowledge, this is the first work applying the neuron-connection level sharing to solve the CVR multi-task learning. This approach dramatically expands the ability to resolve the sharing\&conflict problem in CTR-CVR multi-task learning, which is a paradigm shift compared with the previous coarse-grained sub-network/layer level-sharing methods.

\item We highlight the effectiveness of our approach and challenge it against the single-task model and the classic layer level sharing model, demonstrating relative reductions of 3.78\% and 2.48\% in CVR MSE, respectively. 

\item We deploy our technique in an online video platform with hundreds of millions of users serving main traffic and achieve a significant online performance improvement over current methods, confirming its value in real-world industrial applications.
\end{itemize}


\section{THE PROPOSED APPROACH}
In this section, we introduce our algorithm in detail by initially presenting the typical implementation of the single task CVR and the layer level sharing MTL CVR. Then, we present and discuss the neuron-connection level-sharing method, including theory analysis, tackling the sharing\&conflict issue, and the difference of our method compared to existing works.

\subsection{Single Task \& Layer Level Sharing CVR}
Fig.\ref{fig.main2} presents a typical single task CVR and a layer level sharing CVR architecture. The left part illustrates DLRM \cite{naumov2019deep}, which for this paper represents the single-task baseline. The first layer is an embedding layer for input features, while the second layer is a feature-cross layer that initially involves a Kronecker product of the embedding layer, and then it concatenates the original embedding layer to form the final input features. The following layers are multi-layer perceptrons (MLPs) comprising of a sequence of fully connected (FC) layers, while the output is obtained via a sigmoid operation. The right part of Fig.\ref{fig.main2} illustrates a classic layer level-sharing implementation. The CVR task shares the embedding layer with the CTR task, and thus the fundamental representation of the features (user id, user age, gender, item id, item category, etc.) learned by the CTR task, can be fully shared to compute the CVR score by feed-forward propagation within the network. The remaining layers are not shared.

As mentioned above, ESMM series algorithms and MMOE series algorithms are different implementations of the layer-level sharing method. It should be emphasized that the connection-level CVR sharing is a totally different research paradigm at the granularity of sharing compared with the existing layer-level CVR sharing methods, which means that we don't have to compare it with all the mentioned layer-level sharing algorithm variants.
For simplicity, we choose this most commonly used architecture as the multi-task CVR baseline model for this paper. 
\begin{figure*}[htbp]
\vspace{-15pt}
\centering 
\includegraphics[width=0.9\textwidth,height=0.27\textheight]{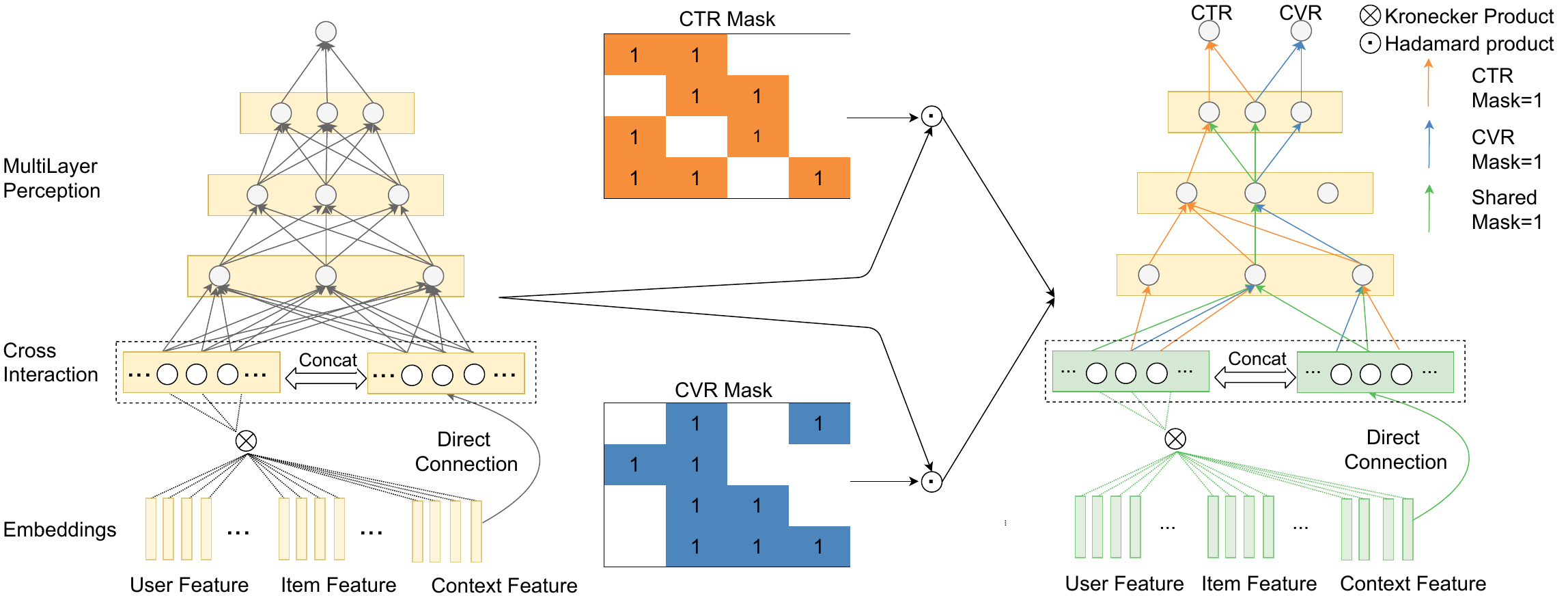} 
\vspace{-10pt}
\caption{Architecture of neuron-connection level sharing CVR modeling. The final model (right) is generated through Hadamard product between the original network (left) and masks (middle). CTR mask=1 means it is only used by CTR, similar to CVR mask=1; shared mask=1 means it is shared by CTR and CVR. The bottom feature embeddings are fully shared. } 
\label{fig.main3} 
\vspace{-12pt}
\end{figure*}

\algnewcommand{\LineComment}[1]{\State \(\triangleright\) #1}
\renewcommand{\algorithmicrequire}{\textbf{Input:}} 
\renewcommand{\algorithmicensure}{\textbf{Output:}}
\begin{algorithm}[htb]
\caption{ CTR/CVR neuron-connection level sharing }
\label{alg:Framwork}
\begin{algorithmic}[1]
    \Require
      Shared network $sNET$; masks $masks[n\_tasks]$; number of pruning $n\_pruning$; number of batches in an epoch of all task mixed samples $n\_batches$; $\odot$ means Hadamard multiplication.
    \Ensure
      CTR model and CVR model;
    \LineComment{Step 1: Warmup}
        \State Train $sNET$ with mixed CVR and CTR samples to get weights $sNET\_init$.
    \LineComment{Step 2: Generate CTR/CVR subnetworks}
        \For{$task\_id$ in [CTR,CVR]}
            \State Initialize $masks[task\_id][0]$ with all values set to 1.
        	\For{$i=1$ to $n\_pruning$}
        		\State a) Train $sNET$ for an epoch of the $task\_id$ task with its samples after $sNET=sNET \odot masks[task\_id][i-1]$;
        		\State b) Let x equal to the qth quantile of the absolute value of the weights from $sNET$, where q is a preset 
        		\Statex \qquad \qquad parameter. Then update the values of $masks[task\_id][i-1]$ by setting all the values less than x to be 0 
        		\Statex \qquad \qquad to obtain $mask[task\_id][i]$;
        		\State c) Reset the network weight of $sNET$ to $sNET\_init$.
        	\EndFor
            \State Select the best $i$ of $masks[task\_id][i]$ according to the performance on the validation set as 
            \Statex $\ \ \ \ \ \ masks[task\_id][best\_i]$. 
        \EndFor
    \LineComment{Step 3: Training weights of CTR/CVR subnetworks in Parallel}

        \State Reinitialize $sNET$ parameters as $sNET\_init$.
        \For{$batch=1$ to $n\_batches$}
       		\State a) Obtain the $task\_id$ of the current sample $batch$,e.g., CTR;
        	\State b)  Update $sNET$ weights feeding $masks[task\_id][best\_i]$;
        	\State c) Train $sNET$ to calculate loss and gradient;
            \State d) Gradient update of weights.
        \EndFor
        \State $CTR\_model=sNET \odot masks[CTR][best\_i]$.
        \State $CVR\_model=sNET \odot masks[CVR][best\_i]$.
\end{algorithmic}
\end{algorithm}

\subsection{Neuron-Connection Level Sharing CVR}
Unlike the layer-level sharing model that specifies which layers to share, in our work, we design mask networks respectively for CTR and CVR, allowing tasks to flexibly learn which neural connections should be activated either for both tasks or solely for a specific task. The proposed method is illustrated in Fig.\ref{fig.main3} and presented in detail in Algorithm \ref{alg:Framwork}.

\textbf{Representation Sharing by Generating Sub-networks}. In MTL, Closely related tasks tend to extract similar sub-nets so they can use similar parts of weights, while loosely related or unrelated tasks tend to extract sub-nets that are different in a wide range \cite{frankle2018the,2020Learning}. Particularly, the inductive bias \cite{baxter2000model} customized to the task is embedded into the subnet structure to some extent. Ideally, tasks with similar inductive bias should be assigned similar parts of parameters.
In practice (Algorithm \ref{alg:Framwork}), we first use samples of all tasks to train a mixed network, i.e., a single model with samples from both CTR and CVR tasks. Then, each task uses its own sample to execute the "mask generation" to remove the neuron-connections (parameters) unimportant for the task itself and constructs a specific sub-network, as is shown in Step 2 of Algorithm 1. This process aims at determining the overlapping part of CTR's sub-network and CVR's sub-network, which represents the same inductive bias held by these two tasks. The unshared part of the network represents each task's own specific inductive bias. 

\textbf{Conflict Reduction Discussion}. As presented in Algorithm \ref{alg:Framwork} the process starts with learning a fully shared base network named $sNET$. Each task first learns its own matrix $mask\in\left\{0,1\right\}$. Then the sub-network of the current task is obtained through $sNET \odot mask$, where $\odot$ denotes the Hadamard product. As shown in Fig.\ref{fig.main3}, CTR and CVR occupy the yellow part and the blue part, respectively, and share the green part in the final model. The co-occupied green part indicates that CTR and CVR will simultaneously share the representation of these parameters, and CTR and CVR still hold their specific representation through the yellow part and the blue part, respectively. This sharing representation and task-specific representation exactly respond to the typical multi-task learning paradigm, which tries to maximize the represent sharing while reducing the conflict. Since the masks/sub-networks are learned automatically, we avoid analyzing which part of the network should be handled with representation sharing and which part with conflict solving techniques \cite{ma2018entire,wen2020entire,zhao2019recommending,ma2019snr,tang2020progressive}. These techniques include manual experience or a large number of hyper-parameter experiments which is hard to guarantee the benefits. 

 \textbf{Sharing Granularity in CVR Multi-task Learning.} Compared with existing literature, the major innovation of this work is the granularity of sharing. MMoE series and ESMM series are different implementations of general layer level-sharing MTL that share bottom layers but present a restricted share at the MLP layers. Indeed, the granularity of sharing is presented at the layers or sub-layers, and these preset sharing layers rely on effective human guidance. Given that this is very different from the proposed neuron-connection level sharing, we pose a quite different research direction. In fact, the layer/sub-network level sharing can be formulated into the framework of our neuron-connection level sharing. If there is no overlap between the mask matrix of CTR and the mask matrix of CVR, that is, the value of the same position will not be 1 simultaneously in both masks, then the neuron-connection level sharing (with the embedding layer still fully shared) is equivalent to the baseline layer level sharing. In this way, connection-level sharing provides more flexible and subtle mechanism for representation sharing and conflict reduction. 

\subsection{Loss Function}
As a multi-stage algorithm (Algorithm 1), we apply stage-specific strategies to train our model. 
At the stage of generating subnetworks, 
we use Eq.(\ref{loss_single}) to extract subnetworks for each task. 
$M_k$ represents the Mask of task $k$, $W_s$ represents all the network weights of both tasks.
$\delta_{k}^i \in \{0,1\}$ indicates whether sample $i$ lies in the sample space of task $k$.
At the stage of joint loss optimization for MTL, we consider to union the sample space of all tasks as a whole training set with Eq.(\ref{equation_loss}), 
where $\theta$ denotes all the network parameters of the MTL model. 
Here $K=\{CTR, CVR\}$ represents the task set. 
Each task ignores samples out of its own sample space when calculating its corresponding loss $L_k$.
Notably, $\lambda_k \in \{0,1\}$ indicates whether the current batch of samples belongs to task $k$.
$\omega_k$ is the loss weight for task $k$. 
At the warmup stage, we also follow Eq.(\ref{equation_loss}).  
 \begin{equation}  
\setlength{\belowdisplayskip}{1pt}
  \begin{split}
  L_k(x;M_k, W_{s}) 
  &=\frac{1}{\sum_i \delta_{k}^i}  \sum_i \delta_{k}^i loss_k(\hat{y}_k(x;M_k\odot W_{s}),y_k)   
\end{split}   
 \label{loss_single}     
\end{equation}

\begin{equation} 
\setlength{\abovedisplayskip}{1pt}
\begin{split} L(x;\theta) 
   &=\sum_{k=1}^K \lambda_k \ast \omega_k  \ast L_{k} 
\end{split}     
\label{equation_loss} 
 \end{equation}

\section{Experiments}
In this section, both offline and online experiments are performed on a large-scale industry online video platform with tens of millions of videos for hundreds of millions of users to evaluate our proposed method. 
\subsection{Experimental Setup}
 \textbf{Datasets.}  In our trials, we employ traffic logs of 9 consecutive days from the video recommender system for offline training and evaluation, for there is no public video recommendation datasets with both clicks and conversions are found.
In the offline experiment, samples in the first 8 days are used for training, and the last day for testing. The dataset used during trials is about 10\% proportion of the online real traffic log, which is extracted through random sampling. Table \ref{table_dataset} summarizes the dataset's statistics.

\textbf{Metric.} We evaluate the performance of the proposed method and the competitor models on two tasks, namely CVR and CTR. Although our main goal is improving CVR, evaluating the CTR performance reveals the effectiveness of our method on this auxiliary data-dense task. The latter is valuable as typical multi-task learning solutions for CVR often sacrifice the performance of CTR. 
\textbf{(1) offline metric}. Mean Squared Error (MSE, the smaller the better) and Area under the ROC curve (AUC, the bigger the better) are adopted as the metrics of CVR and CTR, respectively. \textbf{(2) online metric}. We use the view time as the metric. the predicted view time is given by Eq.(\ref{equationranksocore}), where pCTR and pCVR represents the predicted CTR and CVR, and $\alpha, \beta, \gamma$ are hyper-parameters. We can strengthen the effect of a factor in the final view time prediction by increasing its index. In the online ranking stage of the recommendation system, we will select the highest k videos based on the predicted view time and expose them to the visiting user.
\begin{equation} 
\setlength{\abovedisplayskip}{2pt}
\setlength{\belowdisplayskip}{2pt}
\label{equationranksocore} 
\begin{split} 
 rankscore&=pCTR^\alpha \ast pCVR^\beta \ast video\_length^\gamma \\  
 &=\underset{}{\underset{}{p(y=1\vert x})^\alpha} \ast \underset{}{\underset{}{p{(z\vert y=1)}^\beta}} \ast video\_length^\gamma \end{split},  
  \end{equation}
where $x,y,z$ represent video impression, click, view, respectively.

\textbf{Hyper-Parameter Setting.} 
(1) network setup. We use the Adam optimizer and the Xavier initialization for all methods. The learning rate is 0.0001 and the batch size is 4096 for both models. The network size is 1680*512*256*128 for all competitors except from the layer level-sharing model, which adds a 512*256*128 tower for the CTR task. 
(2) offline training setup. \textbf{$\omega_{ctr}, \omega_{cvr}$} represent for the weight of CTR and CVR loss respectively (Eq.(\ref {equation_loss})). We set $\omega_{ctr}= 0.7$ and  $\omega_{cvr}=0.3$ for both multi-task models. This hyper-parameters is obtained through manual tuning. 
(3) online experiment setup. For simplicity and fairness, we set $\alpha=1, \beta=1, \gamma=1$ in  (Eq.(\ref{equationranksocore})) for all the online experiments independently of the model evaluated. 

 \subsection{Experiment Results} 
 \textbf{Competitors.} The competitor methods are: 
(1) \textbf{Single\_Task} are two classic single-task models for CTR and CVR separately (Fig.\ref{fig.main2}). 
(2) \textbf{Layer\_Share} is the classic layer-level sharing model that shares bottom layers and has the task-specific head layers (Fig.\ref{fig.main2}). 
(3) \textbf{Connection\_Share} is the proposed method as presented in Algorithm \ref{alg:Framwork} (Fig.\ref{fig.main3}). 


\textbf{Offine Experiment}. 
 The offline experiment results are illustrated in Table \ref{table2}.  For an industrial recommender system, a relative 1\% reduction in CVR MSE and a 0.001 absolute gain of CTR AUC are both remarkable and can acquire significant online performance improvement. Specifically, The MSE of the neuron-connection level-sharing CVR is reduced relatively by 3.38\% and 2.49\% compared to the single-task model and the layer sharing model, respectively. This improvement validates the effectiveness of our approach for CVR modeling. The neuron-connection sharing CTR achieves an absolute 0.003 AUC gain and a minor gain of 0.0007 over the two baselines respectively, which is less significant compared to that of CVR. This is expected because CTR samples make up the majority of the entire sample set, and therefore CTR is harder to get a performance gain from MTL.

 
\textbf{Online A/B Testing}. 
Accordingly, the online A/B testing results are illustrated in Table \ref{table3}. We observed the online performance in the video platform for 7 consecutive days, which is immediately after the date of the previous offline training samples. The first three experiments show that the neuron-connection level sharing CVR outperforms the single task and the layer level sharing models. Specifically, experiment 3) shows a \textbf{1.5\%} increase over the single-task model in online view time. Experiment 4) explores the online benefit of our method in the CTR model. The improvement of the CVR metric is probably because the neuron-connection level sharing CTR model is trained with the neuron-connection level sharing CVR model, and may be helpful to the latter. 

\begin{table}
\vspace{-10pt}
\small
\caption{Statistics of the Experimental Dataset}  
\vspace{-10pt}
\label{table_dataset}  
\begin{tabular}{cccccc}
\hline
Dataset       & \#User & \#Video & \#Impression & \#Click & \#Conversion \\ \hline
Tencent Video & 10M    & 11M     & 297M         & 121M    & 49M          \\ \hline
\end{tabular}
\label{tab_dataset}
\vspace{-5pt}
\end{table}

\begin{table}[tbp]
\vspace{-5pt}
\caption{Offline Comparison of Different Models}  
\vspace{-10pt}
\label{table2}  
\setlength{\tabcolsep}{1.8mm}{
\begin{tabular}{lcccc}
\hline
\multicolumn{1}{c}{\multirow{2}{*}{Model}} & MSE         &AUC                       &\multicolumn{2}{c}{MTL Gain}      \\ 
                       & CVR         &CTR                         &      CVR      &  CTR      \\
   \hline
1) Single\_Task      & 0.13688     & 0.78572                    & - & -              \\ 
2) Layer\_Share        & 0.13563          & 0.78808                &   +0.91\%        & +0.30\%  \\
3) Connection\_Share   & \textbf{0.13226} & \textbf{0.78874}      &\textbf{+3.38\%}          & \textbf{+0.38\%} \\ 
\hline
\end{tabular}}

\vspace{-5pt}
\end{table}

\begin{table}[htbp]
  \small
  \caption{Online A/B Test Comparison of Different Models}  
  \vspace{-10pt}
  
  \label{table3}  
  \begin{threeparttable}
  \setlength{\tabcolsep}{1.1mm}{
  \begin{tabular}{lcll}
  \hline
  Rankscore                              & View Seconds  & CVR   & CTR                       \\ \hline
  1)single\_ctr*single\_cvr*length         & 491.3    & 54.9\% & 24.7\%  \\
  2)single\_ctr*layer\_cvr*length              & 481.6                    & 55.0\% & \textbf{24.8\%}                     \\
  3)single\_ctr*connection\_cvr*length     & \textbf{498.6} & 55.1\% & 24.7\%                     \\
  4)connction\_ctr*connection\_cvr*length & 497.1     & \textbf{55.3\%} & \textbf{24.8}\%                     \\ \hline
  \end{tabular}}
  
  \label{tab3}
  \begin{tablenotes}
          \footnotesize
          \item[1] single\_ctr/cvr, layer\_ctr/cvr, connction\_ctr/cvr respectively represent the Single\_Task, Layer\_Share, Connection\_Share models in Table \ref{table2}.
  \end{tablenotes}
  \end{threeparttable}
  \vspace{-10pt}
  \end{table}

\section{CONCLUSIONS AND FUTURE WORK}
In this paper, we present an innovative architecture to optimize the CVR modeling using a neuron-connection level sharing MTL scheme. 
Compared to current literature regarding coarse-grained sub-network/layer level sharing, our fine-grained sharing tackle the difficult sharing\&conflict problem in CTR-CVR MTL by automatically learning which connection weights need to be shared. Experiments on both offline and online industrial video recommender systems demonstrate the superiority. 
Future work should address multi-tasks' personalized sharing of the embedding layer and explore the method in more data-sparse tasks, such as like rate, comment rate.

\bibliography{LT4REC}

\end{document}